# Performance Analysis of HR Portal Domain Components Extraction

N Md Jubair Basha[#], Salman Abdul Moiz [*], A. A. Moiz Qyser [#]

[#]IT  Department, MJCET, Osmania University,Hyderabad, India.

[*]Member, IEEE, Hyderabad, India.

*Abstract*—**Extraction of components pertaining to a particular domain not only reduces the cost but also helps in delivering a quality product. However, the advantages of the Component Level Interaction's (CLI's) are not clearly presented. In the first part of the paper the design of HR Portal application is described. Later the results are simulated using the Netbeans Profiler tool which exposes and highlights the performance characteristics of component based system pertaining to HR domain.**

## I.  INTRODUCTION

Component-based software engineering is an approach to software development that relies on software reuse. It is emerged from the failure of object-oriented development to support effective reuse. Single object classes are too detailed and specific. Building large software intensive-systems is complex. Coping with the views of different stakeholders and balancing the forces upon the systems have inherent engineering challenges. It is indeed a great challenge to maintain individual detailed snowflake contributions to the great avalanche of the entire building of complex software.

This paper presents modelling of HR portal application and the results are simulated to show the performance characteristics of extraction of components.

The remaining part of this paper is organized as follows: Section-II presents different types of components and their representations, Section-III presents domain engineering concepts, section –IV presents the modelling of HR portal application, section –V presents the simulated results using Netbeans Profiler and section-VI concludes the paper.

## II. COMPONENT AND COMPONENT-LEVEL REPRESENTATIONS

The term *component* has been used repeatedly, yet a definitive description of the term is elusive. Brown and Wallnau [1] suggest the following possibilities:
• *Component*— a non-trivial, nearly independent, and replaceable part of a system that fulfils a clear function in the context of a well-defined architecture.
• *Run-time software component*—a dynamic bindable package of one or more programs managed as a unit and accessed through documented interfaces that can be discovered in run time.
• *Software component*—a unit of composition with contractually specified and explicit context dependencies only.
• *Business component*—the software implementation of an "autonomous" business concept or business process.

A Business Component is a web service as a component which is a collection of both business and data with the features of modularity, extensibility and reusability [2].

In addition to COTS components, the CBSE process yields:
• *Qualified components*—assessed by software engineers to ensure that not only functionality, but performance, reliability, usability, and other quality factors conform to the requirements of the system or product to be built.
• *Adapted components*—adapted to modify (also called *mask* or *wrap*) [1] unwanted or undesirable characteristics.
• *Assembled components*—integrated into an architectural style and interconnected with an appropriate infrastructure that allows the components to be coordinated and managed effectively.
• *Updated components*—replacing existing software as new versions of components become available.

Booch el at..[3] defined the component definition as a replaceable part of a system that conforms to and provides the realization of a set of interfaces. Component is the part of the software that can be reused to construct other software or system. The good example for of the software reuse is the component/architecture technique [13].

Component Level Interactions (CLIs), captures component communication, that can be recorded as the program executes. There are two different ways in which CLIs can be represented and the information is recorded, viz., application and system levels. It is important to be able to collect the different component interaction representations by giving examples of the areas in which the information can be used.

CLIs can be represented in different ways. Typical examples include call traces [4], call graphs [5], runtime paths [6], [7], and calling context trees (CCTs) [8]. The different representations contain many levels of information, have different space requirements, and, as such, are suitable for different tasks. Jerding et al. [4] in the past had discussed a spectrum of representations that can be used for the purpose of tracing system execution. They explained the representations ranging from the most accurate and space inefficient to the least accurate and most efficient. Dynamic call traces are said to be the most accurate representation, while at the same time being the most space inefficient. Call graphs in contrast are the least accurate representation but the most space efficient [4], [8].

Young et al..[9], proposed a method for extracting components using the connector-based model. Existing methods use a connector to describe the interactions between components, but regard a connector as a tool for extracting components. And, concentrate on reducing time and effort during components modelling with connector model.

## III. INTRODUCTION TO DOMAIN ENGINEERING

It has been stated that "reuse will disappear, not by elimination, but by integration" into the fabric of software engineering practice [10]. As greater emphasis is placed on reuse, some believe that domain engineering will become as important as software engineering over the next decade. Software Reuse can be improved by identifying objects and operations for a class of similar systems, i.e., for a certain domain. In the context of software engineering domains are application areas. The examples of domains include airline reservation systems, software development tools, user interfaces and financial applications. The scope of a domain can be choosen arbitrarily, either broad for example, banking, or as narrow as text editing. The broad domains are built on top of several narrow domains. For example, user interface domain may be considered as subdomain of the airline reservation systems domain( and several others)[11][12].

Domain Engineering can be defined as by identifying candidate domains and performing domain analysis and domain implementation which includes both application engineering and component engineering.

Domain Analysis is a continuing process of creating and maintaining the reuse infrastructure in a certain domain. The main objective of domain analysis is to make the whole information readily available When the relevant domain is achieved then the relevant components has to be extracted from the repository rather than building the new components from the scratch for the particular domain.

Domain Analysis mainly focuses on reusability of analysis and design, but not code. This can be achieved by building common architectures, generic models or specialized languages that additionally improve the software development process in the specific problem area of the domain.

A vertical domain is a specific class of systems. A horizontal domain contains general software parts being used across multiple vertical domains. Mathematical functions libraries container classes and Unix tools are the examples of horizontal reuse.

The purpose of domain engineering is to identify objects and operations of a class of similar systems in a particular problem domain.

In the context of domain analysis, primarily for the component identification the components can be distinguished with the domain as follows:

**General-purpose components** can be used in various applications of different domains (horizontal reuse).

**Domain-specific component**s are more specific and can be used in various applications of one domain(vertical reuse)

**Product-specific component**s are very specific and custom-built for a certain application; they are not reusable or only to a very small extent.

[14] provided a new software architecture of the component based MIS an example of extracting domain component objects and Subject Document Based Domain Analysis maodel(SDBDA). is proposed for domain analysis and applied an application platform of the tobacco industry.

### 1V. DESIGN OF HR PORTAL DOMAIN

The design of the HR Portal application approach is meant for the recruitment system for a company. The design of the HR Portal application consists of 18 components. Among them there are 8 web components(JSPs and Servlets) and 5 business tier components, all are stateless session beans. The 5 DAO's(Data Access Objects) which will be communicating with the web and business tier components to the database. The application allows candidates to register, log in and browse, view the interview results etc., as well as the details of the employees after the recruitment.

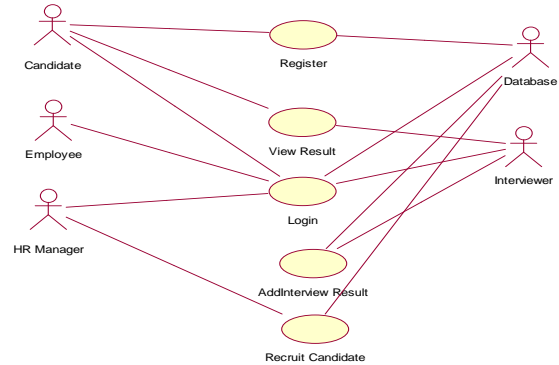

Figure 1: Use Case Diagram for HR Portal

In Figure 1, the analysis model consists of 5 use cases and 5 actors are used. The actors used are Candidate, Employee, HR Manager, Interviewer and Database. Register, View Result, Login, Add Interview Result and Recruit Candidate are the use cases. The set of activities of each use case is described below:

**Register Use Case:** It represents the activity of registration of the new candidates.

**Log in Use Case:** The registered users can make use of the functionality of Log in. Similarly, Employee, HR Manager and Interviewer have access right of Log in Use case.

**Add Interview result Use case:** This use case represents the activity of adding the interview result given by the Interviewer which will be further stored in the database.

**Recruit Candidate Use case:** It represents the activity of recruiting the candidate. Only the HR Manager can invoke this use case

**View Result Use case:** It represents the activity of allowing the candidates to view the interview result which is categorized into level1, level2 and level3.

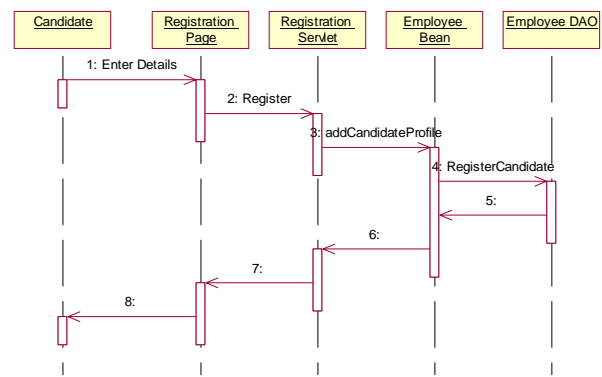

Figure 2: Candidate Registration Sequence Diagram for HR Portal

The Figure 2, consists of a sequence diagram for Candidate Registration with the five objects candidate, registrationpage, registrationServlet, EmployeeBean, EmployeeDAO which interact to realize the analysis model.Since EnterDetails is a prerequisite for all the interactions like register, addCandidateProfile, registerCandidate, the pre-condition for realizing these use cases is enter details.

access objects.Only HR Manager can only recruit the candidate.In turn HRDAO is accessed back to the HRProcessBean, HRProcessServlet, recruitmentpage and HRManager.

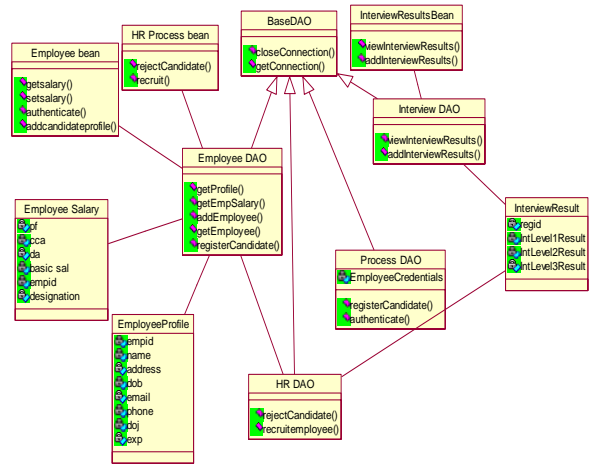

Figure 5:Business-Tier Class Diagram for HR Portal

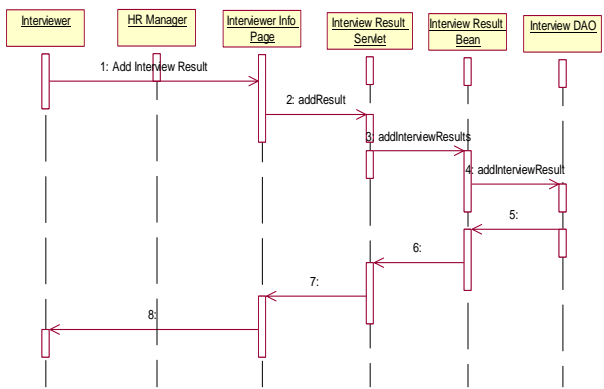

Figure 3: Add Interview Result Sequence Diagram for HR Portal

In Figure 3, there are six objects Interviewer, HR Manager, InterviewInfoPage, InterviewResultServlet, InterviewResultBean and InterviewDAO that interact to realize the use cases.Since addInterviewResult is a prerequisite for all the interactions like addResult, addInterviewResults to the database with the data access objects.

The pre-condition for realizing the use cases ia Add Interview result. Both HR Manager and Interviewer can view the InterviewInfoPage. InterviewerDAO, InterviewResultsBean, InterviewResultsServlet, InterviewInfoPage, HR Manager and Interviewer are interacted to realize domain analysis model.

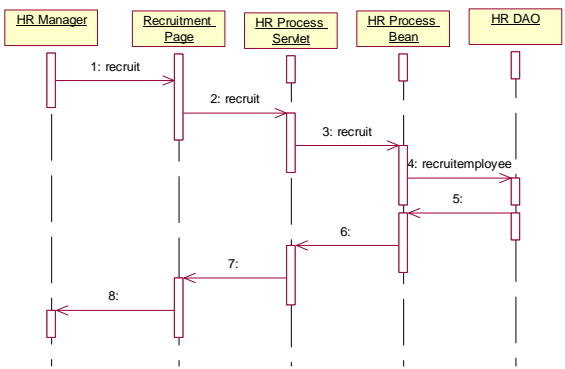

Figure 4: Recruit Sequence Diagram for HR Portal

The sequence diagram of Figure 4, the five objects HR Manager, recruitmentPage, HRProcessServlet, HRProcessBean and HRDAO interact with each other.Since recruit is a prerequisite for the interactions like recruit, and recruitEmployee to the database with the data

The design of Figure 5 is about the business tier class diagram in which the relevant classes are showned with its methods. The classes are EmployeeBean, HRProcess, BaseDAO, EmployeeDAO, InterviewDAO, HRDAO, ProcessDAO, EmployeeProfile, EmpSalary, InterviewResult, InterviewResultsBean. Among these classes there are three stateless beans and five are the data access objects. EmployeeBean, InterviewResultsBean, HRProcessBean are the three stateless bean classes.

Under the EmployeeBean class, the getSalary(), authenticate(), addCandidatePofile() methods are designed. The getSalary() will give the salary of the employee which is reflected from the EmpSalary class and further connected to the EmployeeDAO object of the database. The authenticate() method will give registered user can sign in the Login page whose user name are available in the database. The addCandidateProfile() is designed when the new registered user added to the BaseDAO. InterviewRessultsBean class consists of methods viewInterviewResults() and addInterviewResults(). The viewInterviewResults() methods is designed to check the result of the candidate who is pass or fail. The addInterviewresults() method is designed to add the Interviewed Results of the candidate depends upon the InterviewLevels. HRProcessBean class consists of methods rejectCandidate() and recruit().Depending upon the InterviewResltsBean, the recruit() and rejectCandidate() methods are designed to display the result. Thus, all the stateless beans in the HR Portal are designed.

The BaseDAO, EmployeeDAO, InterviewDAO, HRDAO, ProcessDAO are designed as the data access objects to the database. Among these objects, BaseDAO is the root for the other objects such as InterviewDAO, HRDAO, EmployeeDAO, ProcessDAO. BaseDAO class contains getConnection() and closeConnection() methods. The purpose of designing the getConnection() method is to connect to the database and closeConnection() method is to close the database connections. The EmployeeDAO consists of methods as getProfile(), getEmpSalary(),

addEmployee(), getEmployee(), registerCandidate(). The design of these methods is based on the employee details. The getProfile() method is designed to give the Employee details such as empid, name, address, date of birth, email, phone, date of joining, experience. The getEmpSalary() method is to give the employee salary as pf, cca, hra, da, basicSal, empId, designation. The addEmployee() method is designed to add the employee to the database. The getEmployee() method is designed to give the Employee details such as empid, name, address, date of birth, email, phone, date of joining, experience. The registerCandidate() method is designed to register the candidate with his registration details which were specified in RegistrationServlet.The database tables are Employee Profile, Employee Salary, Interview Result and Employee Credentials.

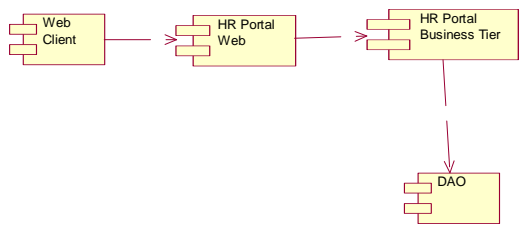

Figure 5: Component Diagram for HR Portal

The component diagram for HR portal is depicted in figure 5.The system is designed in such a way that the client can interact with the web tier and business tier and can connect to the Data Access Object(DAO).
The web-tier consists of the JSP's and Servlets.The Business tier consists of the EJB's.The DAO's consists of the classes with its objects communicating to the database.The web-tier components are HttpServlet, HRProcessServlet, Login Servlet, InterviewResultServlet and RegistrationServlet classes.The Business-tier components are EmployeeBean, InterviewResultsBean, HRProcessBean are the three stateless bean classes.The DAO components are BaseDAO, EmployeeDAO, InterviewDAO, HRDAO, ProcessDAO classes.

## V. RESULTS OF HR PORTAL APPLICATION DOMAIN

For each application, it have a non instrumented version, and a version that has been instrumented with the Netbeans profiler[15]. For the Netbeans profiler, have set a filter to instrument the application components and the EJB container calls (i.e., the com.sun.ejb.* and javax.ejb.*).

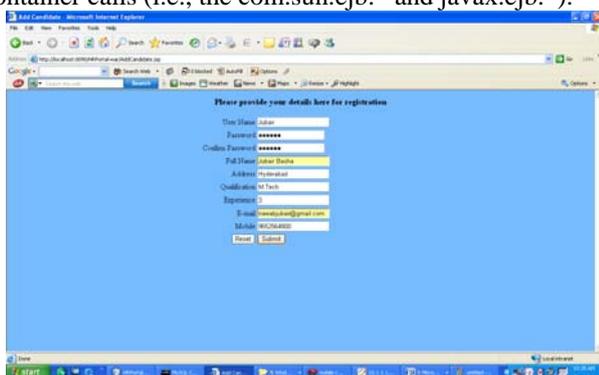

Figure 6: Registration details page of HR Portal Application

In this paper, the HR Portal application is used to for the interaction of the components, Figure 6 shows the registration details of a user which collects the components for the invocation purpose.The registration details includes the Username, Password, Onfirm Password, Full Name, Address, Qualification, Experience, E-mail, Mobile text fileds , submit and reset buttons in the Registration Web page. Figure 7 provides the Monitoring, Performance Analysis and Memory details of the HR Portal application.The analysis performance can be done to the entire application or on parts. The profile of the project and subproject classes can be filtered.
Netbeans collects CLIs in the form of CCTs. The CCTs give a summary view of the component interactions and the related performance characteristics helps in knowing the system performance. Figure 8 shows a CCT that gives a summary of the calls corresponding to the HR Portal "Log in" user action of Login.jsp and Register.jsp web pages. The performance metrics relate to snapshots that were taken, i.e., when 20 and 50 users were using the system, respectively. The columns in the diagram correspond to the name of method called, the total time spent in this method for all invocations (including time spent in calls made by this method) and the total number of calls.

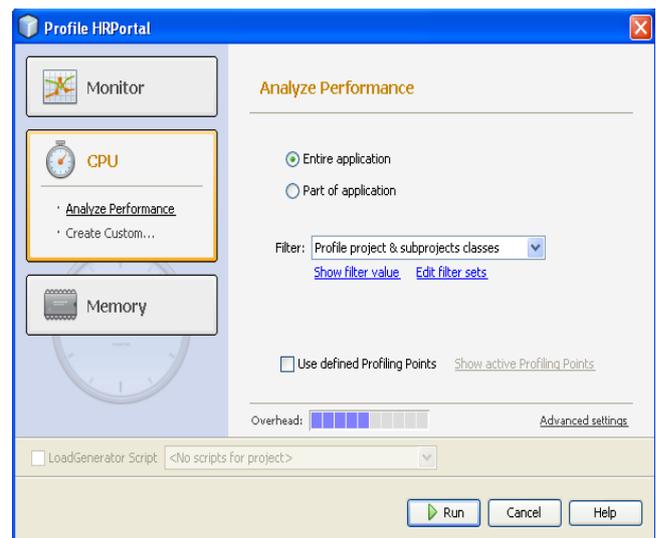

Figure 7: Performance Analysis of HR Portal Application

Figure 8: CCTs of Components in the HRPortal Application

Figure 8 shows CCTs for HR Portal Application for the 20-user snapshot. The total amount of time spent in the 10 invocations of the EmployeeÐAO. authenticateEmployee() method is 15.2 ms (i.e., an average of 1.52 ms per invocation). The CCT contains application-level calls and middleware calls related to the EJB container. The information in Fig 6 shows that even when 20 users are using the system the average response time of the application-level calls related to the "log in" user action is similar to that when a single user is in the system in fact shows an invocation of this method. The BaseDAO.getConnection() method is 1267ms (i.e., an average of 24.92ms per invocation). As well as EmployeeDAO.addCandidateProfile() method is taking 946ms and 47.3ms per invocation. The Login_jsp._jspservice() is taking 54.6ms for 20 users and making 2.184ms per invocation. In order to identify the components that had the biggest impact on the overall response time at 20 users, we further analyzed the CCT. Figure 8 shows a number of the container calls that are invoked prior to an invocation of the com.mycompany.hr.dao.BaseDAO.init() with 20 users in the system.

As shown in Figure 8 , a large proportion of the time is spent mainly in the Web container and EJB container calls. As the CCTs contain a summary of method call information, they can be easily used to identify method hot spots. Because of the level of our instrumentation filter settings (we chose to instrument the EJB container in detail only), we could not investigate further where time was being spent in the Web container. The percentage of utilization of the every component is also showned. For BaseDAO.getConnection() method, the percentage of utilization is 41.4%, EmployeeDAO.addCandidateProfile() method 30.9%,EmployeeDAO addEmployeeCredentials() method 20.48%, EmployeeDAO. authenticateEmployee() method is 2.8% and so on.

Thus dynamically Component level interaction extraction is achieved by using the NetBeans Profiler tool.

The Telemetric view of the number of Threads/Loaded Classes are with the Maximum and Current Threads is also provided .

It is believed that the higher levels of overhead associated with the Netbeans profiler in comparison to other profiling tools that can be attributed to the fact that the Netbeans profiler obtains more detailed information, collecting both system- and application-level data. Furthermore, the Netbeans profiler(by default) also collects basic heap, garbage collection(see Figure 9), and threading information, which is not obtained using other tools. Thus, when there is an increase in the number of users in the system, the overhead related to obtaining CLIs increases. The information related heap collection, garbage collection, and thread information is also depicted.

## VI. CONCLUSION

Component systems for varied domains help in effective reuse of the product or system help in developing systems within estimated time and budget. To know the advantages of the component extraction, Netbeans profiler is used to study and analyse the performance of component based systems. Study on component interactions for other domains such as e-governance, e-Banking etc., can be done by making a repository of specific components which might be reused efficiently. One of the research issues is the mechanism to secure the identified components at the middleware level. Design and development of new component extraction tools is needed to enhance the power of reusability. Another research issue is to reduce the effort of product development.

### ACKNOWLEDGEMENTS

The work was partly supported by the R & D Cell of Muffakham Jah College of Engineering & Technology, Hyderabad, India. The authors would like to thank to all the people from Industry and Academia for their active support.

### REFERENCES

[1]  Brown, A.W. and K.C. Wallnau, "Engineering of Component Based Systems ",*Component-Based Software Engineering*, IEEE Computer Society Press, 1996, pp. 7–15.
[2]  Feng Jiao, Liping Wang " A Business Component Model for Domain-Specific Software" 978-1-4244-4507-3/09, IEEE, 2009.
[3]  Grady Booch, James Rambaugh, Ivar Jacobson "The Unified Modelling Language User Guide", Second Edition, Fourth Impression 2008, Pearson Education.
[4]  S.L. Graham, P.B. Kessler, and M.K. McKusick, "GPROF: A Call Graph Execution Profiler," Proc. SIGPLAN Symp. Compiler Construction, 1982.
[5]  M. Chen, E. Kiciman, A. Accardi, A. Fox, and E. Brewer, "Using Runtime Paths for Macro Analysis," Proc. Ninth Workshop Hot Topics in Operating Systems, 2003.
[6]  T. Parsons, "Automatic Detection of Performance Design and Deployment Antipatterns in Component Based Enterprise Systems," PhD dissertation, Univ. College Dublin, 2007.
[7]  G. Ammons, T. Ball, and J.R. Larus, "Exploiting Hardware Performance Counters with Flow and Context Sensitive Profiling," Proc. ACM SIGPLAN Conf. Programming Language Design and Implementation, 1997.
[8]  M. Trofin and J. Murphy, "Static Verification of Component Composition in Contextual Composition Frameworks," Int'l J.Software Tools for Technology Transfer, Jan. 2008.
[9]  Young Ran Yu, Soo Dong Kim, Dong Kwan Kim "Connector Modeling Method for Component", IEEE, 1999.

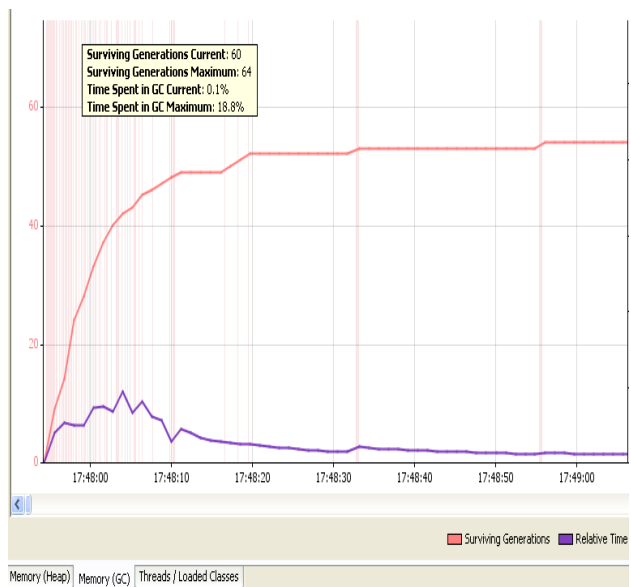

Figure 9: Garbage Collector Graph of HR Portal application in Memory Analysis

[10]   Tracz, W., "Third International Conference on Software Reuse—Summary," ACM Software Engineering Notes, vol. 20, no. 2, April 1995, pp. 21–22.
         Ruben Prieto-Diaz, "Domain Analysis for reusability" in COMPSAC 87, pages 225-234, Tokya, Japan, 1987.
[11]   Ruben Prieto-Diaz, "Domain Analysis: An Introduction", ACM SIGSOFT Software Engineering Notes, 15(2):47-54, April 1990.
[12]   Li Yu cai, Wang Ling, "Re-integrating of Architecture/Component Orientedd General Model of Sub-domain", IEEE, 2005
[13]   Geng gangyong, Zhong cuihao, Chen wei  "A Domain-Specific Software Architecture", IEEE, 1997.
[14]   The Netbeans Profiler, http://profiler.netbeans.org, 2008

## ABOUT THE AUTHORS

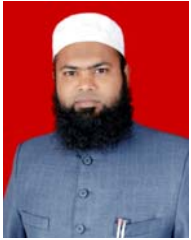

**N Md Jubair Basha** received his B.Tech. (IT) and M.Tech (IT) from JNTUH, Hyderabad. He is presently working as Assistant Professor in Department of Information Technology, Muffakham Jah College of Engineering and Technology, Hyderabad, India. His research interest includes Software Reusability, Data Mining and Mobile Computing. He is a life member of Computer Society of India. You can reach him at nawabjubair@gmail.com.

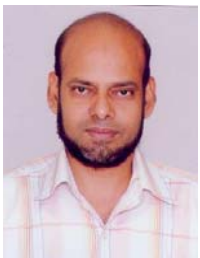

**Dr. Salman Abdul Moiz** received B.Sc (Electronics) from Osmania University, MCA from Osmania University, M.Tech(CSE) from Osmania University, M.Phil (CS) from Madurai Kamaraj University and Ph.D(CSE) from Osmania University.  He worked as Research Scientist at Centre for Development of Advanced Computing, Bangalore. He has published 27 papers in various National/International Conferences and Journals. His areas of interests include Mobile databases, Software Process Improvements; Component based software development & Disaster Recovery. He is an active member of IEEE, IETE and CSI.

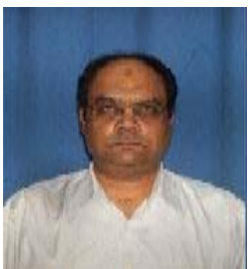

**Dr. Ahmed Abdul Moiz Qyser** received his B.E. (CSE) from Osmania University, M.Tech. (Software Engineering) from JNTU, Hyderabad, and Ph.D. from Osmania University. His research focus is Software Process Models and Metrics for SMEs. He is presently working as Professor & Head in Department of Information Technology, Muffakham Jah College of Engineering and Technology, Hyderabad, India. He is also a visiting Professor to the industry where he teaches Software Engineering and its related areas. He is the author of several research papers in the area of Software Engineering. He is an active member of ACM, CSI and ISTE.